\documentclass[prl,twocolumn,showpacs]{revtex4}
\usepackage{graphicx}
\usepackage{times}
\usepackage{bm}
\usepackage{amsmath, verbatim}

\def\be{\begin{equation}}
\def\ee{\end{equation}}
\def\bea{\begin{eqnarray}}
\def\eea{\end{eqnarray}}
\def\bse{\begin{subequations}}
\def\ese{\end{subequations}}

\def\be{\begin{eqnarray}}
\def\ee{\end{eqnarray}}

\begin{document}

\title{A generic new platform for topological quantum computation using semiconductor heterostructures}
\author{Jay D. Sau$^1$}
\author{Roman M. Lutchyn$^1$}
\author{Sumanta Tewari$^{1, 2}$}
\author{S. Das Sarma$^1$}
\affiliation{$^1$Condensed Matter Theory Center and Joint Quantum Institute, Department of Physics, University of
Maryland, College Park, Maryland 20742-4111, USA\\
$^2$Department of Physics and Astronomy, Clemson University, Clemson, SC
29634}

\begin{abstract}
We show that a film of a semiconductor in which $s$-wave superconductivity and a Zeeman splitting are induced by proximity effect,
supports zero-energy Majorana fermion modes in the ordinary vortex excitations. Since time reversal symmetry is explicitly broken, the edge of the film constitutes a chiral Majorana wire. The heterostructure we propose --  a semiconducting thin film sandwiched between an $s$-wave superconductor and a magnetic insulator -- is a generic
 system which can be used as the platform for topological quantum computation by virtue of the existence of non-Abelian Majorana fermions.
\end{abstract}

\maketitle


\paragraph{Introduction.}

In two spatial dimensions, where permutation and exchange are not necessarily equivalent, particles can have quantum statistics which are strikingly different from the familiar
statistics of bosons and fermions. In situations where the many body ground state wave-function is a linear combination of states from a degenerate subspace,
a pairwise exchange of the particle coordinates can unitarily \emph{rotate} the ground state wave-function in the degenerate subspace.
In this case, the exchange statistics is given by a multi-dimensional unitary matrix representation (as opposed to just a phase factor)
 of the 2D braid group, and the statistics is non-Abelian~\cite{nayak_RevModPhys'08}.
It has been proposed that such a system, where the ground state degeneracy is protected by a gap from local perturbations, can be used as a fault-tolerant platform for topological quantum computation (TQC)~\cite{dassarma_prl'05}.

Recently, the $\nu = 5/2$ FQH state at high magnetic fields and at low temperature has been proposed as a topological qubit ~\cite{dassarma_prl'05}. This theoretical conjecture, however, awaits experimental verification~\cite{stern_prl'06, bonderson_prl'06}. An equivalent system, in which the ordered state is in the same universality class as the $5/2$ FQH
state, is the spin-less (spin-polarized) $p_x+ip_y$ superconductor/superfluid~\cite{Read_prb'00}. In a finite magnetic field, a vortex excitation in such a superconductor traps a single, non-degenerate, zero-energy bound state. The key to non-Abelian statistics is that the second-quantized operator for this zero-energy state is self-hermitian, $\gamma^{\dagger}=\gamma$, rendering $\gamma$ a Majorana fermion operator. If the constituent fermions have spin, the spin-degeneracy of the zero energy excitation spoils the non-Abelian statistics, because the mutual statistics of the vortex-fermion composites becomes trivial. To circumvent this problem in a realistic, spinful, $p_x+ip_y$ superconductor such as strontium ruthenate, it has been proposed that the requisite excitations are the exotic half-quantum vortices, which can be thought of as ordinary vortex excitations in only one of the spin sector in the condensate~\cite{ivanov_prl'01}.

Even though quenching the spin-degeneracy by either the application of a magnetic field ~\cite{DasSarma_PRB'06} or by using spin-less atomic systems \cite{tewari_prl'2007} is possible in principle, it is practically very difficult. Therefore, it is desirable to have systems whose most natural excitations themselves follow non-Abelian statistics \emph{in spite of} the fermions carrying, as they do in a realistic system, a spin quantum number. The recent proposal by Fu and Kane~\cite{fu_prl'08} points out one such system -- the surface of a strong TI in proximity to an $s$-wave superconductor -- which supports a non-degenerate Majorana fermion excitation in the core of an ordinary vortex. In this paper, we propose a simple generic TQC platform by showing that it is possible to replace the
 TI with a regular semiconductor film with spin-orbit coupling, provided the time-reversal symmetry is broken by proximity of the film
to a magnetic insulator. The three ingredients of non-Abelian statistics -- spin-orbit coupling, $s$-wave superconductivity, and Zeeman splitting --
  are experimentally known to occur in many solid state materials. It is encouraging that the $s$-wave
proximity effect has already been demonstrated in 2D InAs heterostructures which additionally also have a 
substantial spin-orbit coupling~\cite{merkt}.  Thus, the structure we propose is one of the simplest to realize
  non-Abelian Majorana fermions in the solid-state context.

\paragraph{Theoretical Model.}

\begin{figure}
\centering
\includegraphics[width=0.75\linewidth,angle=0]{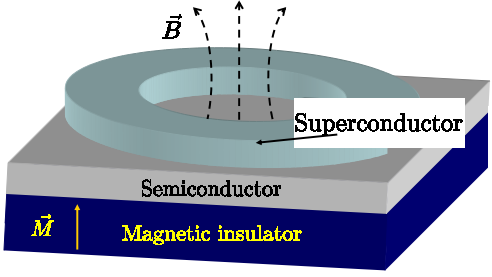}
\caption{(Color online) Schematic picture of the proposed heterostructure exhibiting Majorana zero-energy bound state inside an ordinary vortex.}\label{fig:structure}
\end{figure}

The single-particle effective Hamiltonian $H_0$ for the conduction band of a spin-orbit coupled semiconductor in contact with a magnetic insulator is given by (we set $\hbar=1$ henceforth)
\begin{align}\label{eq:H0}
H_0\!=\! \frac{p^2}{2 m^*}\!-\!\mu\!+\!V_z \sigma_z\!+\!\alpha (\vec \sigma \!\times \! \vec p)\!\cdot\! \hat{z}.
\end{align}
Here, $m^*$, $V_z$ and $\mu$ are the conduction-band effective mass of an electron, effective Zeeman coupling induced
by proximity to a magnetic insulator (we neglect the direct coupling of the electrons with the magnetic field from the magnetic insulator), and chemical potential, respectively.  The coefficient
 $\alpha$ describes the strength of the Rashba spin-orbit coupling and $\sigma_{\alpha}$ are the Pauli matrices. Despite the similarity in the spin-orbit-coupling terms, $H_0$ and the Hamiltonian for the TI surface in Ref.~[\onlinecite{fu_prl'08}] differ by the existence of a spin-diagonal kinetic energy term in Eq.~(\ref{eq:H0}). Because of the spin-diagonal kinetic energy, there are in general two spin-orbit-split Fermi surfaces in the present system, in contrast to the surface of a TI in which
an odd number of bands cross the Fermi level ~\cite{fu_prl'08}.      
In Eq.~(\ref{eq:H0}), for out-of-plane Zeeman coupling such that $|V_z| > |\mu|$, a single band crosses the Fermi level. Thus, analogous to a strong TI surface (but arising from qualitatively different physics), the system
 has a single Fermi surface, which is suggestive of non-Abelian
topological order if $s$-wave superconductivity is induced in the film. We show below that this is indeed the case by analyzing the Bogoliubov de Gennes (BdG) equations for a vortex in the superconductor in the
 heterostructure shown in Fig.~\ref{fig:structure}.


The proximity-induced superconductivity in the semiconductor can be described by the Hamiltonian,
\begin{equation}
\hat{H}_{p}=\int d\mathbf{r}\,\{\Delta_0(\mathbf{r})\hat{c}^{\dagger}_{\uparrow}(\mathbf{r})\hat{c}^{\dagger}_{\downarrow}(\mathbf{r})+\rm H.c\},
\end{equation}
where $\hat{c}_{\sigma}^{\dagger}(\mathbf{r})$ are the creation operators for electrons with spin $\sigma$ and $\Delta_0(\mathbf{r})$ is the proximity-induced gap. The corresponding BdG equations written in Nambu space become,
\begin{equation}
\left(\begin{array}{cc}H_0&\Delta_0(\mathbf{r})\\\Delta_0^*(\mathbf{r})&-\sigma_y H_0^* \sigma_y\end{array}\right)\Psi(\mathbf{r})=E\Psi(\mathbf{r}),
\end{equation}
where $\Psi(\mathbf{r})$ is the wave function in the Nambu spinor basis, $\Psi(\mathbf{r})=(u_{\uparrow}(\mathbf{r}),u_{\downarrow}(\mathbf{r}),v_{\downarrow}(\mathbf{r}),-v_{\uparrow}(\mathbf{r}))^T$. Using the solutions of the BdG equations, one can define Bogoliubov quasiparticle operators as $\hat{\gamma}^\dagger=\int d\mathbf{r}\,\sum_{\sigma}u_{\sigma}(\mathbf{r})\hat{c}_{\sigma}^{\dagger}(\mathbf{r})+v_{\sigma}(\mathbf{r})\hat{c}_{\sigma}(\mathbf{r})$. The bulk excitation spectrum of the BdG equations with $\Delta(r)=\Delta_0$ has a gap for non-vanishing spin-orbit coupling.

\paragraph{BdG equations for a vortex.}
We now consider the vortex in the heterostructure shown in
Fig.\ref{fig:structure}, and take the vortex-like
configuration of the order parameter: $\Delta_0(r,\theta)=\Delta_0(r)e^{\imath \theta}$. Because of the rotational symmetry, the BdG equations can be decoupled into angular momentum channels indexed by $m$ with the corresponding spinor wave-function,
\begin{equation}
\!\Psi_m(r,\!\theta)\!=\!e^{\imath m \theta}\left(u_{\uparrow}(r),u_{\downarrow}(r)e^{i\theta},\!v_{\downarrow}(r)e^{ -i\theta},-v_{\uparrow}(r)\right)^T\!.
\end{equation}
Because of the particle-hole symmetry of the BdG equations, if $\Psi(\mathbf{r})$ is a solution with energy
$E$ then $\imath\sigma_y\tau_y \Psi^*(\mathbf{r})$ is also a solution at energy $-E$. Here $\tau_y$ is defined to be the Pauli matrix in Nambu spinor space. In particular, zero-energy solutions of the BdG equations must come in pairs, $\Psi(\mathbf{r})$ and
$\imath\sigma_y\tau_y \Psi^*(\mathbf{r})$, unless these two wave-functions refer to the same state. Thus, a zero-energy solution in an angular momentum channel $m$ is always paired with another zero-energy solution in the channel $-m$ and, therefore, can be non-degenerate only if it corresponds to the $m=0$ angular momentum channel.

The radial BdG equations describing the zero-energy state in the $m=0$ channel can be written as
\begin{align}
&\left(\begin{array}{cc}H_0&\Delta_0(r)\\\Delta_0(r)&-\sigma_y H_0^* \sigma_y\end{array}\right)\Psi(r)=0,\\
\!H_0\!=\!&\!\left(\!\begin{array}{cc}\!-\eta (\partial_r^2\!+\!\frac{1}{r}\partial_r)\!+\!V_z\!-\!\mu\! & \alpha (\partial_r\!+\!\frac{1}{r} )\!\\\\ - \alpha \partial_r  \! & \! \eta (-\partial_r^2\!-\!\frac{1}{r}\partial_r\!+\!\frac{1}{r^2})\!-\!V_z\!-\!\mu\! \end{array}\right)\nonumber
\end{align}
with $\eta=\frac{1}{2 m^*}$. Since the BdG matrix is real, there are two solutions  $\Psi(r)$ and $\Psi^*(r)$ with $E=0$.
 Furthermore, it follows from the particle-hole symmetry of the BdG equations that $\imath\sigma_y\tau_y \Psi(r)$ is also a solution. Thus, any non-degenerate  $E=0$ solution must satisfy the property $\imath\sigma_y\tau_y \Psi(r)=\imath\lambda \Psi(r)$.
Moreover, because $(\imath\sigma_y\tau_y)^2=-1$, the possible values of $\lambda$ are $\lambda=\pm 1$.
The value of $\lambda$ sets a constraint on the spin-degree of freedom of $\Psi(r)$, such that
$v_{\uparrow}(r)=\lambda u_{\uparrow}(r)$ and $u_{\downarrow}(r)=\lambda v_{\downarrow}(r)$. This allows one to eliminate
the spin degrees of freedom in $\Psi(r)$ and define a reduced spinor $\Psi_0(r)=(u_{\uparrow}(r),u_{\downarrow}(r))^T$. The corresponding reduced BdG equations take the form of a $2\times 2$ matrix differential equation:
\begin{align}\label{eq:zeroenergy}
\!&\!\left(\!\begin{array}{cc}\!\!-\!\eta (\partial_r^2\!+\!\frac{1}{r}\partial_r)\!+\!V_z\!-\!\mu\!&\! \lambda\Delta(r)\!+\!\alpha (\partial_r\!+\!\frac{1}{r} )\\\\ -\lambda \Delta(r)\!-\!\alpha \partial_r  \!&\! -\!\eta (\partial_r^2\!+\!\frac{1}{r}\partial_r\!-\!\frac{1}{r^2}\!)\!-\!V_z\!-\!\mu\! \end{array}\!\right)\!\!\Psi_0(r)\!=\!0.
\end{align}
We now approximate the radial dependence of $\Delta_0(r)$ by $\Delta_0(r)=0$ for $r<R$ and $\Delta_0(r)=\Delta_0$ for $r\geq R$. In view of the  stability of the putative Majorana zero-energy solution to local changes in the Hamiltonian \cite{Read_prb'00}, such an approximation can be made without loss of generality.  For $r<R$ the analytical solution of Eq.~\eqref{eq:zeroenergy} is given by $\Psi_0(r)=\left(u_{\uparrow} J_0 (z r ), u_{\downarrow} J_1(z r)\right)^T$ with the constraint
\begin{align}\label{eq:z1}
\left(\begin{array}{cc}\eta z^2+V_z-\mu & z\alpha  \\ \alpha z  & \eta z^2 -V_z-\mu \end{array}\right)\left(\begin{array}{c} u_{\uparrow} \\ u_{\downarrow}\end{array}\right)=0.
\end{align}
Here $J_n(r)$ are the Bessel functions of the first kind. The characteristic equation for $z$ is
\begin{equation}\label{eq:z11}
(\eta z^2-\mu)^2-V_z^2-\alpha^2 z^2=0.
\end{equation}
In the case $\mu > V_z$ the roots of Eq.\eqref{eq:z11} are real: $z_{1,2}=\pm \sqrt{w_1}$ and $z_{3,4}=\pm \sqrt{w_2}$ with $w_{1,2}>0$ being the solutions for $z^2$. In the opposite limit, $0 < \mu < V_z$, there are two real solutions $z_{1,2}=\pm \sqrt{|w_1|}$ and two imaginary solutions $z_{3,4}=\pm i \sqrt{|w_2|}$. Since the Bessel functions are symmetric, we find two solutions which are well-behaved at the origin: $\Psi_1(r)=(u_{\uparrow} J_0(z_1 r),u_{\downarrow} J_1(z_1 r))^T$ and $\Psi_2(r)=(u_{\uparrow} J_0(z_3 r),u_{\downarrow} J_1(z_3 r))^T$. Therefore, the full solution at $r<R$ is $\Psi^{<}_0(r)=c_1\Psi_1(r)+c_2\Psi_2(r)$.

\begin{figure}
\centering
\includegraphics[width=0.99\linewidth,angle=0]{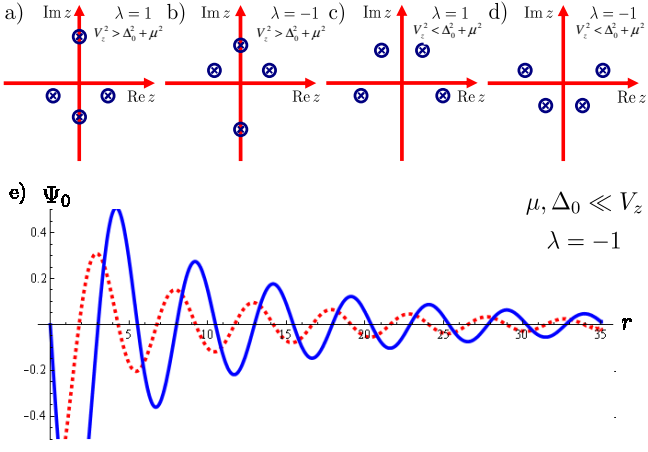}
\caption{(Color online) Upper panel: complex roots of the Eq.~\eqref{eq:z2} for different values of $\mu$ and $\lambda$ with $\Delta_0\neq 0$. Lower panel: Numerical solution for the Majorana zero-energy state $\Psi_0(r)=(u(r),v(r))^T$ for $\lambda=-1$ and $\mu<V_z$. The dashed (red) and solid (blue) lines correspond to $u(r)$ and $v(r)$, respectively. Here we used the following parameters: $\eta=\alpha=V_z=1$, $\mu=0$, $\Delta_0=0.1$ and $R=1$. The boundary conditions used are $\Psi_0(0)=(1,0)^T$ and $\Psi_0(r=40)=(0,0)^T$. }\label{fig:roots}
\end{figure}

At large distances $r>R$, where $\Delta_0(r)=\Delta_0$, the solution to Eq.~\eqref{eq:zeroenergy} is complicated. Nevertheless, one can write the solution as a series expansion in $1/r$:
\begin{equation}
\Psi_0(r)=\frac{e^{\imath z r}}{\sqrt r}\sum_{n=0,1,2...}\frac{a_n}{r^n}
\end{equation}
where $a_n$ are the corresponding spinors. The zeroth order coefficient  $a_0$ satisfies the following equation:
\begin{equation}\label{eq:z2}
\left(\begin{array}{cc}\eta z^2+V_z-\mu & \lambda\Delta_0+\imath z\alpha  \\ -\lambda \Delta_0-\imath z \alpha   & \eta z^2 -V_z-\mu \end{array}\right)a_0=0.
\end{equation}
 The higher order coefficients $a_n$ can be calculated from $a_0$ using a set of recursion relations \cite{Jaydeep_unpublished}. The characteristic equation for Eq.~(\ref{eq:z2}) has 4 complex roots for $z$, which are shown in Fig~\ref{fig:roots}. Physical solutions of Eq.~\eqref{eq:zeroenergy} at $r>R$, $\Psi_0^{>}(r)=\sum_{n>2}c_n \Psi_n(r)$, require that ${\rm Im}[z_n]>0$. (Here $\Psi_n(r)$ is the solution corresponding to the eigenvalue $z_n$) Thus, for $(\mu^2+\Delta_0^2) >V_z^2$, there are two solutions for $\lambda=\pm 1$.
 On the other hand, for $(\mu^2+\Delta_0^2) <V_z^2$ there are three solutions for $\lambda=-1$ and only one for  $\lambda=1$.

In order to obtain a unique solution for the zero-energy state, the wavefunctions $\Psi_0^{>}(r)$ and $\Psi_0^{<}(r)$ should satisfy boundary conditions at $r=R$.
Since we are matching 2-component wave-functions and their derivatives at $r=R$, the continuity of $\Psi_0(R)$ and $\Psi'_0(R)$ leads to 4 independent equations. One additional constraint comes from the normalization of the wavefunction in all space. Thus, there are five independent constraints for the coefficients $c_n$. A unique zero-energy solution exists if the number of unknown coefficients $c_n$ is five, which is the case for $(\mu^2+\Delta_0^2) <V_z^2$ and $\lambda=-1$. In this case, the wavefunctions $\Psi_0^{<}(r)=\sum_{n=1,2}c_n\Psi_n(r)$ and $\Psi_0^{>}(r)=\sum_{n=3,4,5}c_n\Psi_n(r)$ have 2 and 3 unknown coefficients, respectively. In all other cases the number of unknowns $c_n$ is smaller than five, and thus, as we have checked explicitly, solutions for the zero-energy eigenfunction do not exist. From these arguments one can conclude that, for $(\mu^2+\Delta_0^2) <V_z^2$, an ordinary vortex in the superconducting condensate contains a unique non-degenerate $E=0$ solution in the $m=0$  angular momentum channel.
The numerical solution for the zero-energy state is shown in Fig.~\ref{fig:roots}e. It is straightforward to check that the zero-energy solution corresponds to a self-hermitian second-quantized operator
$\gamma=\gamma^\dag$: it is a Majorana fermion excitation.

We can also consider the special case with $\eta=0, V_z=0$ in the above equations, which
describes the recent proposal for TQC \cite{fu_prl'08} using zero-energy Majorana bound states
at vortices on the interface of a TI and an $s$-wave
 superconductor. In this case, we find a single solution for $r<R$ and a pair of independent solutions for $r>R$. Since the BdG differential equation is now
 only first order,
we need only match the 2-component spinors themselves (derivatives need not match) which yields 3 equations for the 3 coefficients.
This leads to a unique Majarana fermion solution at the vortex, which is consistent with Ref.~[\onlinecite{fu_prl'08}]. Interestingly, in contrast to our Hamiltonian for $\eta>0$, the condition for the existence of a Majorana fermion for
 $\eta=0$ is given by $V_z^2<(\Delta_0^2+\mu^2)$.
The model considered in  Ref.~[\onlinecite{fu_prl'08}] and our present system have a similar order parameter structure.
 In both cases the order parameter component $\langle c_{\uparrow}(r)c_{\downarrow}(r')\rangle$ has an $s$-wave orbital symmetry while
 the order parameter components $\langle c_{\uparrow}(r)c_{\uparrow}(r')\rangle$ and
 $\langle c_{\downarrow}(r)c_{\downarrow}(r')\rangle$ have
 $p_x+\imath p_y$ and $p_x-\imath p_y$ orbital symmetries, respectively. On the surface of a TI, because of time-reversal
invariance, $|\langle c_{\uparrow}(r)c_{\uparrow}(r')\rangle|=|\langle c_{\downarrow}(r)c_{\downarrow}(r')\rangle|$.
In our system, the ratio of the order parameter components in the two spin sectors is different from 1,
 and  approaches 1 in the limit $\alpha^2/\eta\gg V_z$. In both cases, however, the superconducting pairing potential is $s$-wave, and is induced by proximity effect. Therefore, the superconducting state and the associated non-Abelian topological character are expected to be robust in the presence of finite disorder.

\paragraph{Topological phase transition.}
 We have shown above that a non-degenerate Majorana state exists in a vortex in the superconductor only
in the parameter regime $(\mu^2+\Delta_0^2) < V_z^2$.
This suggests that there must be a quantum phase transition (QPT) separating the
parameter regimes $(\mu^2+\Delta_0^2) <V_z^2$ and $(\mu^2+\Delta_0^2) >V_z^2$, even though the system in both regimes is an $s$-wave superconductor. A non-degenerate zero-energy solution cannot disappear
unless a continuum of energy levels appears around $E=0$. Such a continuum of states at $E=0$ can only appear if the bulk gap closes,
which can be used to define a topological quantum phase transition. In the present system, such a phase transition
can be accessed by varying either the Zeeman splitting or the chemical potential. A similar topological 
quantum phase transition has already been predicted for ultra-cold atoms with vortices in the 
spin-orbit coupling~\cite{fujimotosato}.

\begin{figure}
\centering
\includegraphics[scale=0.3,angle=0]{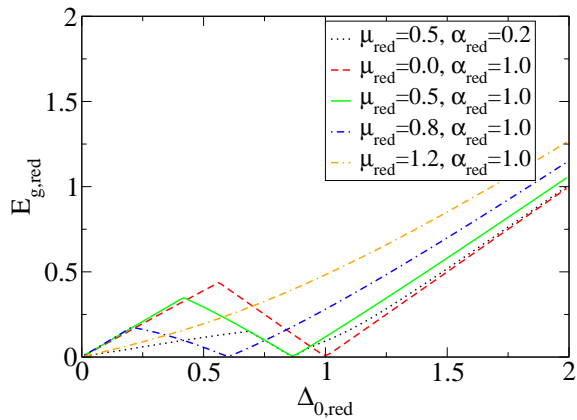}
\caption{(Color online) Quasiparticle gap versus pairing potential for various values of the chemical potential $\mu$. Here $E_{g,\rm red} =E_g/V_z$, $\Delta_{0,\rm red}=\Delta_0/V_z$, $\mu_{\rm red}=\mu/V_z$ and $\alpha_{\rm red}=\alpha/\sqrt{\eta V_z}$. The maximum value of the gap in the topologically non-trivial superconductor, and the corresponding area in the phase diagram, decreases with increasing values of $\mu$. For large negative $\mu$, the system makes a transition to a semiconductor. The
phase to the right of the critical point is the topologically trivial $s$-wave superconductor.}
\label{figure:gap}
\end{figure}

The bulk gap of the present system can be calculated from the bulk excitation spectrum,
\begin{equation}
E^2=V_z^2+\Delta_0^2+\tilde{\epsilon}^2+\alpha^2 k^2\pm 2 \sqrt{V_z^2\Delta_0^2+\tilde{\epsilon}^2(V_z^2+\alpha^2 k^2)}
\end{equation}
where $\tilde{\epsilon}=\eta k^2-\mu$.
As seen in Fig.~\ref{figure:gap}, the excitation gap first increases as a function of $\Delta_0$ (proximity-induced pair-potential) and then decreases and vanishes at
a critical point, $\Delta_{0c}=\sqrt{V_z^2-\mu^2}$, before re-opening and increasing with $\Delta_0$.
The critical point marks the phase transition between a topologically non-trivial (left) and a topologically trivial (right) $s$-wave superconducting phases. The scale of the gap in the topologically non-trivial phase is set by the strength of the spin-orbit coupling $\alpha$ and the position of the critical point. The fact that the phase
on the right-side of the critical point does not support a non-degenerate Majorana mode can be verified by observing that, for these values of
$\Delta_0$, it is possible to reduce $V_z$ such that $|V_z|<|\mu|$  without the gap
 vanishing at any point. This is the phase without Majorana Fermion excitations. In fact, this phase can be reached
 from the conventional $s$-wave superconductor with $V_z=0$ and $\alpha=0$ without crossing a phase transition.

\paragraph{Majorana edge modes and TQC.}
In analogy with Ref.~\cite{fu_prl'08}, we find that an interface
between two superconductor layers, which can be deposited on the semiconductor thin film, supports a pair of zero-energy excitations when the phase difference between the
superconductors is $\pi$. This geometry can be analyzed in a way that closely follows our derivation of the localized state in a vortex in the
$m=0$ channel, since the BdG Hamiltonian can again be reduced to a real matrix. In this case, we find that, in the parameter
regime $(\mu^2+\Delta_0^2) <V_z^2$, there are 3 linearly independent solutions on each side of the interface. Since the number of constraints to be satisfied at the interface (we assume the interface to be of negligible width) remains 5 as before,
 one expects a pair of independent zero-energy solutions. The interface, therefore, constitutes a \emph{non-chiral} Majorana wire,
which can be exploited for braiding in a way completely analogous to Ref.~\cite{fu_prl'08} to perform TQC. 
 Majorana bound states as  well as Majorana  edge modes in our system can be studied experimentally using non-local Andreev reflection~\cite{Nilsson_prl'08} and electrically detected interferometry~\cite{fu_prl'09, akhmerov_prl'09} experiments.

The experimental implementation of this proposal involves a heterostructure  of a magnetic insulator (\emph{e.g.} EuO), a strong spin-orbit coupled semiconductor (\emph{e.g.} InAs) and an s-wave superconductor with a large $T_c$ (\emph{e.g.} Nb). Using these materials, it is possible \cite{Jaydeep_unpublished} to induce an effective superconducting pairing potential $\Delta_0\sim 0.5$ meV and a tunneling-induced effective Zeeman splitting  $V_z \sim 1$ meV. Additionally, the strength of spin-orbit interaction $\alpha$ in InAs heterostructures is electric-field tunable and can be made as large as
$\alpha\approx 50$ meV-\AA~\cite{ingaas_prl}. With these estimates, the quasiparticle gap $E_g$ is of the order of 1 K. Given that the chemical 
potential is gate-tunable and can be of the of the order of $\Delta_0$, we numerically estimate the magnitude of the excitation energy for the bound states in a vortex core of size $\sim 20$ nm to be of the order of $0.1$ K \cite{Jaydeep_unpublished}, which sets the temperature scale for TQC in this system.

\paragraph{Conclusion.} Our proposed TQC platform
should be simpler to implement experimentally than any of the TQC candidates proposed in the literature so far, since it involves a standard heterostructure with a magnetic insulator, a semiconductor film, and an ordinary $s$-wave superconductor. We believe that the proposed scheme provides the most straightforward method for the solid-state realization of non-Abelian Majorana fermions. A significant practical advantage of the proposed TQC scheme is its generic simplicity: it requires neither special samples or materials nor ultra-low temperatures or high magnetic fields.

This work is supported by DARPA-QuEST, JQI-NSF-PFC, and LPS-NSA. We thank M. Sato and S. Fujimoto for discussion. ST acknowledges DOE/EPSCoR Grant \# DE-FG02-04ER-46139 and Clemson University start up funds for support.




\end{document}